\documentclass[aps,prd,twocolumn,superscriptaddress,showpacs]{revtex4}
\usepackage[dvips]{graphicx}
\usepackage{amsmath,amssymb,times}

\newcommand{\bequ}{\begin{equation}}
\newcommand{\eequ}{\end{equation}}
\newcommand{\bea}{\begin{eqnarray}}
\newcommand{\eea}{\end{eqnarray}}

\DeclareSymbolFont{boldletters}{OML}{cmm} {b}{it}
\DeclareSymbolFontAlphabet{\mathbit}{boldletters}
\DeclareMathSymbol{\alpha}{\mathalpha}{letters}{"0B}
\DeclareMathSymbol{\beta}{\mathalpha}{letters}{"0C}
\DeclareMathSymbol{\gamma}{\mathalpha}{letters}{"0D}
\DeclareMathSymbol{\delta}{\mathalpha}{letters}{"0E}
\DeclareMathSymbol{\epsilon}{\mathalpha}{letters}{"0F}
\DeclareMathSymbol{\zeta}{\mathalpha}{letters}{"10}
\DeclareMathSymbol{\eta}{\mathalpha}{letters}{"11}
\DeclareMathSymbol{\theta}{\mathalpha}{letters}{"12}
\DeclareMathSymbol{\iota}{\mathalpha}{letters}{"13}
\DeclareMathSymbol{\kappa}{\mathalpha}{letters}{"14}
\DeclareMathSymbol{\lambda}{\mathalpha}{letters}{"15}
\DeclareMathSymbol{\mu}{\mathalpha}{letters}{"16}
\DeclareMathSymbol{\nu}{\mathalpha}{letters}{"17}
\DeclareMathSymbol{\xi}{\mathalpha}{letters}{"18}
\DeclareMathSymbol{\pi}{\mathalpha}{letters}{"19}
\DeclareMathSymbol{\rho}{\mathalpha}{letters}{"1A}
\DeclareMathSymbol{\sigma}{\mathalpha}{letters}{"1B}
\DeclareMathSymbol{\tau}{\mathalpha}{letters}{"1C}
\DeclareMathSymbol{\upsilon}{\mathalpha}{letters}{"1D}
\DeclareMathSymbol{\phi}{\mathalpha}{letters}{"1E}
\DeclareMathSymbol{\chi}{\mathalpha}{letters}{"1F}
\DeclareMathSymbol{\psi}{\mathalpha}{letters}{"20}
\DeclareMathSymbol{\omega}{\mathalpha}{letters}{"21}
\DeclareMathSymbol{\varepsilon}{\mathalpha}{letters}{"22}
\DeclareMathSymbol{\vartheta}{\mathalpha}{letters}{"23}
\DeclareMathSymbol{\varpi}{\mathalpha}{letters}{"24}
\DeclareMathSymbol{\varrho}{\mathalpha}{letters}{"25}
\DeclareMathSymbol{\varsigma}{\mathalpha}{letters}{"26}
\DeclareMathSymbol{\varphi}{\mathalpha}{letters}{"27}
\DeclareMathSymbol{\Gamma}{\mathalpha}{letters}{"00}
\DeclareMathSymbol{\Delta}{\mathalpha}{letters}{"01}
\DeclareMathSymbol{\Theta}{\mathalpha}{letters}{"02}
\DeclareMathSymbol{\Lambda}{\mathalpha}{letters}{"03}
\DeclareMathSymbol{\Xi}{\mathalpha}{letters}{"04}
\DeclareMathSymbol{\Pi}{\mathalpha}{letters}{"05}
\DeclareMathSymbol{\Sigma}{\mathalpha}{letters}{"06}
\DeclareMathSymbol{\Upsilon}{\mathalpha}{letters}{"07}
\DeclareMathSymbol{\Phi}{\mathalpha}{letters}{"08}
\DeclareMathSymbol{\Psi}{\mathalpha}{letters}{"09}
\DeclareMathSymbol{\Omega}{\mathalpha}{letters}{"0A}

\begin{document}
\title{Differences and similarities between fundamental and adjoint matters in $SU(N)$ gauge theories}

\author{Hiroaki Kouno}
\email[]{kounoh@cc.saga-u.ac.jp}
\affiliation{Department of Physics, Saga University,
             Saga 840-8502, Japan}

\author{Tatsuhiro Misumi}
\email[]{tmisumi@bnl.gov}
\affiliation{Department of Physics, Brookheaven, National Laboratory, Upton, NY 11973}

\author{Kouji Kashiwa}
\email[]{kashiwa@ribf.riken.jp}
\affiliation{RIKEN/BNL, Brookheaven, National Laboratory, Upton, NY 11973}

\author{Takahiro Makiyama}
\email[]{12634019@edu.cc.saga-u.ac.jp}
\affiliation{Department of Physics, Saga University,
             Saga 840-8502, Japan}

\author{\\Takahiro Sasaki}
\email[]{sasaki@phys.kyushu-u.ac.jp}
\affiliation{Department of Physics, Graduate School of Sciences, Kyushu University,
             Fukuoka 812-8581, Japan}

\author{Masanobu Yahiro}
\email[]{yahiro@phys.kyushu-u.ac.jp}
\affiliation{Department of Physics, Graduate School of Sciences, Kyushu University,
             Fukuoka 812-8581, Japan}

\date{\today}

\begin{abstract}
We investigate differences  and similarities between fundamental fermions and adjoint fermions in 
$SU(N)$ gauge theories. 
The gauge theory with fundamental fermions possesses $\mathbb{Z}_{N}$ symmetry only in the limit of infinite fermion mass, whereas the gauge theory with adjoint fermions does have the symmetry for any fermion mass. 
The flavor-dependent  twisted boundary condition (FTBC) is then 
imposed on fundamental fermions so that the theory with fundamental fermions can possess $\mathbb{Z}_N$ symmetry for any fermion mass.
We show similarities between FTBC fundamental fermions and adjoint fermions,  using the Polyakov-loop extended Nambu--Jona-Lasinio (PNJL) model.  
 In the mean-field level, the PNJL model with FTBC fundamental fermions has dynamics similar to the PNJL model with adjoint fermions for the confinement/deconfinement transition related to $\mathbb{Z}_N$ symmetry. 
The chiral property is somewhat different between the two models, but 
there is a simple relation between chiral condensates in the two models. 
As an interesting high-energy phenomenon, a possibility of the gauge symmetry breaking is studied for FTBC fundamental fermions. 
\end{abstract}

\maketitle

\section{Introduction}

Understanding of nonperturbative nature of quantum chromodynamics (QCD) is one of the most important subjects in particle physics. QCD  has $\mathbb{Z}_{N}$ symmetry only in the limit of infinite current quark mass ($m$) and 
chiral symmetry only in the limit of $m=0$. 
In the real world where $m$ is finite, some nonperturbative properties 
are discovered by lattice QCD (LQCD). For finite temperature ($T$), for example, deconfinement and chiral transitions are found to be crossover \cite{YAoki_nature}. 
For finite quark chemical potential ($\mu_q$), however, our understanding of nonperturbative nature of QCD is still far from perfection, since LQCD simulations have the sign problem.

QCD is a $SU(3)$ gauge theory with fundamental fermions. 
In this sense, a $SU(N)$ gauge theory with adjoint fermions~\cite{KL_aQCD} is a QCD-like theory. 
This QCD-like theory is quite interesting, since it has $\mathbb{Z}_{N}$ symmetry for any $m$ and no sign problem for finite $\mu_q$ when the number of flavor $N_{F}$ is even. 
Furthermore, if the gauge theory with adjoint fermions is considered on spacetime  $R^4 \times S^1$, there is a possibility that the theory has the Hosotani mechanism~\cite{Hosotani} 
where the gauge symmetry breaking (GB) is induced by the non-zero vacuum expectation value of 
the gauge component in a compact dimension $S^1$. Actually, the GB is found to occur, when the periodic boundary condition is imposed on adjoint fermions; see Ref.~\cite{Kashiwa_GB} and references therein. 
In the case of four-dimensional gauge theory at finite $T$ where 
the spacetime of the theory is compactified into $R^3 \times S^1$,  
exotic phases such as 
the reconfined phase appear at high $T$ , when adjoint fermions are introduced with the periodic boundary condition~\cite{MO,Cossu_EX,Nishimura_ADJ}. 
These interesting properties of adjoint matter lead us to an important question, how close is adjoint matter to fundamental matter? 

The fermion number is $N \times N_F$ for $N_F$-flavor fundamental fermions and  $N^2-1$ for one-flavor adjoint fermions. 
These numbers are almost identical with each other for $N=N_F \gg 1$, and even in the realistic case 
of $N=N_F =3$ they are still close to each other. 
A difference is that the $SU(N)$ gauge theory with fundamental fermions does not possess $\mathbb{Z}_{N}$ symmetry for  finite $m$. 
This difference can be removed by imposing the flavor-dependent twist boundary condition (FTBC) on 
fundamental fermions~\cite{Kouno_TBC,Sakai_TBC}. 
We refer to fundamental fermions with the FTBC as FTBC fundamental fermions in this paper. 
For zero $T$, FTBC fundamental
fermions yield the same dynamics as ordinary fundamental fermions with  the anti-periodic boundary condition, since the fermion boundary condition does not affect dynamics there. 
In Refs.~\cite{Kouno_TBC,Sakai_TBC}, properties of the $SU(N)$ gauge theory with FTBC fundamental fermions were investigated with 
the Polyakov-loop extended Nambu-Jona-Lasinio (PNJL)  model~\cite{Meisinger,Dumitru,Fukushima,Ratti,Megias,Rossner,
Schaefer,Abuki,Fukushima2,Kashiwa1,McLerran_largeNc,Nishimura_ADJ,Kahara_ADJ,Zhang_ADJ,Sakai,Sasaki-T_Nf3}. 
The PNJL model with FTBC fundamental fermions has $\mathbb{Z}_N$ symmetry for any $m$. 
In the model, the symmetry is preserved at low $T$ but spontaneously broken at high $T$, and the restoration of chiral symmetry is rather slow. 
Similar properties are seen in the $SU(3)$ gauge theory with adjoint fermions. 
These results imply that the $SU(N)$ gauge theory with FTBC fundamental fermions has dynamics similar to 
the $SU(N)$ gauge theory with adjoint fermions and hence the GB takes place also in the former theory. 

In this paper, we show similarities between adjoint matter and FTBC fundamental matter in $SU(N)$ gauge theories, 
particularly for the confinement/deconfinement transition related 
to $\mathbb{Z}_N$ symmetry, using the PNJL model. 
This leads to the important conclusion that 
an essential difference between ordinary fundamental matter and 
adjoint matter is originated in the presence or absence 
of $\mathbb{Z}_{N}$ symmetry. 
Meanwhile, the chiral property is somewhat different between 
adjoint matter and FTBC fundamental matter, 
but we show that there is a simple relation between 
chiral condensates in the two matters. 
As an interesting high-energy phenomenon, a possibility of the GB is also examined for FTBC fundamental fermion and the result 
is compared with that for adjoint fermion. 

This paper is organized as follows. 
In Sec. \ref{fermion}, thermodynamics of a gauge theory with adjoint fermion and a gauge theory with FTBC fundamental fermion are constructed 
in a similar manner.  
In Sec. \ref{sec:PNJL}, 
similarities between FTBC fundamental and adjoint matters 
are numerically analyzed with the PNJL model. 
In Sec. \ref{sec:GB}, 
a possibility of the GB is discussed for FTBC fundamental matter by using the one-loop effective potential  and the result is compared with that for adjoint matter.  
Section \ref{sec:summary} is devoted to a summary.

\section{Adjoint and FTBC fundamental fermions}
\label{fermion}

In this section, 
thermodynamics of 
a $SU(N)$ gauge theory with adjoint fermions and a $SU(N)$ gauge theory with FTBC fundamental fermions~\cite{Kouno_TBC,Sakai_TBC} are constructed in a similar way. 
For later convenience, we start with a $SU(N)$ gauge theory with fermions in the $N\times \bar{N}$ dimensional representation consisting  of 
the  $N^2-1$ dimensional adjoint and the one-dimensional singlet representation. Here we consider a general case of 
$N_{F,adj}$ degenerate flavors, where $N_{F,adj}$ means the number of flavors for adjoint fermions.  
In Euclidean spacetime, the Lagrangian density $L_{adj}$  becomes
\bea
L_{adj}= N_{F,adj}\bar{\Psi}(\gamma_\nu D_\nu^{N\times \bar{N}} +m)\Psi
+{1\over{4g^2}}{F_{\mu\nu}^{a}}^2 
\label{QCD-CC}
\eea
with $D_\nu^{N\times \bar{N}} \equiv \partial_\nu-i(A_\nu+\tilde{A}_\nu) = \partial _\nu -iA_{a,\nu}(t_a-\tilde{t}_a)$. The fermion field $\Psi$ in its color part is described as a direct product of $\psi_c$ and $\tilde{\psi}_c$ ($c=1,2,\cdots,N$), where $\psi_c$ ($\tilde{\psi}_c$) is transformed as the $N$ ($\bar{N}$) dimensional representation and the generator $t_a$ ($\tilde{t}_a$) acts only on $\psi_c$ ($\tilde{\psi}_c$). 
Below we consider a general temporal boundary condition for $\Psi$:
\begin{eqnarray}
\Psi(\tau =\beta,{\bf x})=e^{i\varphi}\Psi(\tau =0,{\bf x})  
\label{BC}
\end{eqnarray}
with Euclidean time $\tau$ and $\beta =1/T$. 
The angle $\varphi$ parameterizes the boundary condition. 
A value of $\varphi =\pi~(0)$ corresponds to the anti-periodic (periodic) boundary condition.

The Lagrangian density (\ref{QCD-CC}) is invariant under the large gauge transformation, 
\begin{eqnarray}
\Psi &\to& \Psi^\prime =U\tilde{U}\Psi,
\nonumber
\\
A_\nu &\to& A_\nu^\prime =UA_{\nu}U^{-1}+i(\partial_\nu U)U^{-1}, 
\nonumber
\\
\tilde{A}_\nu &\to& \tilde{A}_\nu^\prime =\tilde{U}\tilde{A}_{\nu}\tilde{U}^{-1}+i(\partial_\nu \tilde{U})\tilde{U}^{-1} ,
\label{Zntrans_CC}
\end{eqnarray}
where  
\begin{eqnarray}
U(x,\tau)&=&\exp{(i\alpha_at_a)}, 
\label{gauge_element_1}
\\
\tilde{U}(x,\tau )&=&\exp{(-i\alpha_a\tilde{t}_a)},  
\label{gauge_element_2}
\end{eqnarray}
are elements of the SU$(N)$ group characterized by real functions $\alpha_a(x,\tau)$ and satisfy the boundary conditions
\begin{eqnarray}
U (x,\beta )&=&\exp{(-i2\pi k/N)}U(x,0),
\\
\tilde{U}(x,\beta )&=&\exp{(i2\pi k/N)}\tilde{U}(x,0)
\label{Z_N_twist}
\end{eqnarray}
with integer $k$. 
In this transformation, $\tilde{\psi}_c$ is transformed as the conjugate representation of $\psi_c$. 
Thus $\Psi$ belongs to the $N\times \bar{N}$ dimensional representation that is decomposed into the  $N^2-1$ dimensional adjoint 
and the one-dimensional singlet representation.  
This theory is finally reduced to a $SU(N)$ gauge theory with adjoint fermions, since the singlet fermion is  decoupled from the adjoint ones.  
The gauge transformation \eqref{Zntrans_CC} includes the $\mathbb{Z}_N$ transformation. 
The Lagrangian density \eqref{QCD-CC} and the fermion boundary condition \eqref{BC} are 
invariant under the $\mathbb{Z}_N$ transformation, since the $\mathbb{Z}_N$ transformation on $\psi_c$ is canceled out by that on $\tilde{\psi}_c$. 
Hence $\mathbb{Z}_N$ symmetry is exact in this theory.

Now a gauge theory with FTBC fundamental fermions is constructed by replacing the color-dependent fields $\tilde{\psi}_c~(c=1,2,\cdots,N)$ by the flavor-dependent ones $\hat{\psi}_f~(f=1,2,\cdots,N)$ in $\Psi$ and the gauge field $i\tilde{A}_{\nu,a}$ by the flavor-dependent imaginary chemical potential $i\hat{B}_\nu =i\Theta_a\hat{t}_a\delta_{\nu,4}T$ with the $SU(N)$ generators $\hat{t}_a~(a=1,2,\cdots,N^2-1)$ and the unit matrix $\hat{t}_0$ in flavor space and real parameters $\Theta_a~(a=0,1,\cdots,N^2-1)$.   
After the replacement, the new fermion field $\Psi_{fund}$ belongs to the $N$ dimensional fundamental representation in color space. Hence one can 	represent $\Psi_{fund}=(\Psi_1, \cdots,\Psi_N)^T$ with 
fundamental fermions $\Psi_f$ labeled by flavors $f=1, \cdots,N$.  
This replacement thus changes $N\times \bar{N}$ dimensional fermions of 
$N_{F,adj}$ flavors to  fundamental fermions  $\Psi_f$ of 
$N_{F,fund}=N_{F,adj}N$ flavors. 
The Lagrangian density  $L_{fund}$ is then  
\bea
L_{fund}= N_{F,adj}\bar{\Psi}_{fund}(\gamma_\nu D_\nu^N + m)\Psi_{fund}
+{1\over{4g^2}}{F_{\mu\nu}^{a}}^2, 
\label{QCD-CF}
\eea
where $D_\nu^N \equiv \partial_\nu-i(A_\nu +\hat{B}_\nu )$. 
This is nothing but a $SU(N)$ gauge theory with 
$N_{F,fund}$ flavor fundamental fermions with the 
flavor-dependent imaginary chemical potential. 
Although the gauge field $A_{\nu}$ is decoupled from the flavor degrees of freedom, 
one can consider the following baryon-number-color-flavor linked (BCFL) transformation 
\begin{eqnarray}
\Psi_{fund} &\to& \Psi_{fund}^\prime =U\hat{U}\Psi_{fund}
\nonumber\\
A_\nu &\to& A_\nu^\prime =UA_{\nu}U^{-1}+i(\partial_\nu U)U^{-1}  
\label{ZNtrans_CF}
\end{eqnarray}
where 
\begin{eqnarray}
&&U(x,\tau)=\exp{(i\alpha_at_a)}
\label{colorU}
\\
&&\hat{U}(\tau )=\exp{(i2\pi k T\tau /N)}(\hat{U}_{\rm Sf})^k, 
\label{CFL-trans}
\end{eqnarray}
where $U$ is the same as in \eqref{gauge_element_1}, but 
$\hat{U}$ with integer $k$ acts on flavor space through a factor $(\hat{U}_{\rm Sf})_{ff^\prime}=\delta_{f-1,f^\prime}$ for $f\neq 1$ and $(\hat{U}_{\rm Sf})_{ff^\prime}=\delta_{N,f^\prime}$ for $f=1$; 
note that $U_{\rm Sf}$ is the $N\times N$ unitary matrix by which flavor labels are shifted from $f$ to $f-1$ for $f\neq 1$ and from $f$ to $N$ for $f=1$. 
The matrix $U_{\rm Sf}$ in flavor space is equal to $i^sU_{\rm SUf}$ with $s=1$ for even $N$ and $s=0$ for odd $N$ and  
an element $U_{\rm SUf}$ of the flavor $SU(N)$ group. 
The transformation \eqref{ZNtrans_CF}, characterized by the $\mathbb{Z}_N$ transformation parameter $k$, 
thus links baryon number, color and flavor. 

The boundary condition for $\Psi_{fund}$ at $\tau =\beta$ is invariant under the transformation \eqref{ZNtrans_CF}, but the Lagrangian density (\ref{QCD-CF}) is changed into
\bea
L_{fund}= N_{F,adj}\bar{\Psi}_{fund}(\gamma_\nu {{D}_\nu^N}^\prime + m)\Psi_{fund}
+{1\over{4g^2}}{F_{\mu\nu}^{a}}^2, 
\nonumber\\
\label{QCD-CF-2}
\eea
where ${D_\nu^N}^\prime =\partial_\nu -i(A_\nu+\hat{B}_\nu^\prime )$ with $\hat{B}_\nu^\prime=\Theta_a^\prime \hat{t}_a\delta_{\nu,4}T$ and 
\begin{eqnarray}
\Theta^\prime _a\hat{t}_a =(\hat{U}_{\rm Sf}^{-1})^k(\Theta_a\hat{t}_a-{2k\pi \over{N}}\hat{t}_0)(\hat{U}_{\rm Sf})^k. 
\label{theta_transform}
\end{eqnarray}
In general, $L_{fund}$  is not invariant under this transformation. 
However, when 
\begin{eqnarray}
\Theta_a\hat{t}_a&=&{\rm diag}(\theta_1,\theta_2,\cdots,\theta_N)
\nonumber\\
&=&{\rm diag}(0,2\pi/N,\cdots, 
\nonumber\\
&& 2(l-1)/\pi/N,\cdots,2\pi (N-1))~~~ 
\label{FTBC}
\end{eqnarray}
with $l=1,2,\cdots, N$,  the transformed imaginary chemical potential 
$\Theta^\prime_a\hat{t}_a$ becomes 
\begin{eqnarray}
\Theta^\prime _a\hat{t}_a&=&(\theta_1^\prime,\theta_2^\prime,\cdots,\theta_N^\prime )
\nonumber\\
&=&(U_{\rm Sf}^{-1})^k{\rm diag}(-2\pi k/N,2\pi (1-k)/N,\cdots, 
\nonumber\\
&& 2\pi (l-k-1)/N,\cdots,2\pi (N-k-1))(U_{\rm Sf})^k
\nonumber\\
&\cong&
{\rm diag}(0,2\pi/N,\cdots, 
\nonumber\\
&& 2(l-1)/\pi/N,\cdots,2\pi (N-1))
\nonumber\\
&=&(\theta_1,\theta_2,\cdots,\theta_N)
=\Theta_a\hat{t}_a, 
\label{FTBC-2}
\end{eqnarray}
where $\theta_f^\prime \cong \theta_f$ means $\theta_f^\prime/(2\pi)=\theta_f/(2\pi)~({\rm mod}~ 1)$; 
note that $\theta_f$ have a trivial periodicity of $2\pi$ 
in the QCD partition function.  
$\mathbb{Z}_N$ symmetry is thus exact in 
a $SU(N)$ gauge theory with fundamental fermions 
of $N_{F, fund}$ degenerate flavors, when 
the flavor-dependent imaginary chemical potential of \eqref{FTBC} is  introduced.      
Obviously, the flavor-dependent imaginary chemical potential 
breaks flavor symmetry partially. 
In the confinement phase, flavor symmetry is recovered from the breaking~\cite{Kouno_TBC,Sakai_TBC}, as shown explicitly 
in Sec. \ref{sec:PNJL}.

 When the fermion fields $\Psi_f$ are transformed as 
\begin{eqnarray}
\Psi_f \to \exp{(i\theta_fT\tau )}\Psi_f
\label{transform_1}
\end{eqnarray}
for $f=1,2,\cdots,N$, the Lagrangian (\ref{QCD-CF}) with the imaginary chemical potential (\ref{FTBC}) is changed into a new Lagrangian density with no imaginary chemical potential,
\bea
L_{fund}&=& N_{adj}\bar{\Psi}_{fund}(\gamma_\nu (\partial_\mu -iA_\nu) + m)\Psi_{fund}
\nonumber \\
&~~&+{1\over{4g^2}}{F_{\mu\nu}^{a}}^2, 
\label{QCD-CF-3}
\eea
with the boundary condition 
\begin{eqnarray}
\Psi_f(\tau =\beta,{\bf x})=e^{i(\varphi -\theta_f)}\Psi_f(\tau =0,{\bf x}). 
\label{FTBC_B}
\end{eqnarray}
The boundary condition \eqref{FTBC_B} with the twist angles $\theta_f$ of \eqref{FTBC} is 
the  flavor-dependent boundary condition (FTBC) and 
fundamental fermions satisfying the FTBC are FTBC fundamental fermions. 
As mentioned above, the $SU(N)$ gauge theory with FTBC fundamental fermions has $\mathbb{Z}_N$ symmetry just as the $SU(N)$ gauge theory with adjoint fermions. 
The fermion number, i.e., the product of the color and flavor numbers, is $N_{F,adj}N^2$ for the former theory 
and  $N_{F,adj}(N^2-1)$ for the latter one. 
These numbers are close to each other even for $N=3$. 
Thus there is a possibility that the two theories have similar dynamics to each other. 
This will be tested in  the following two sections; 
low energy dynamics is analyzed in Sec. \ref{sec:PNJL}, 
while the GB as a nontrivial high-energy phenomenon is investigated in Sec. \ref{sec:GB}.

The adjoint representation is a real representation under the gauge transformation and hence the system has no sign problem even at real $\mu_{q}$, when $N_{F,adj}$  is even. 
The fundamental representation with the FTBC, meanwhile, 
is not real and therefore the system has a sign problem at real $\mu_{q}$. 
However, if we restrict the color gauge transformation $U$ on its center group $\mathbb{Z}_N$ in the color-flavor linked transformation (\ref{ZNtrans_CF}), 
the fermion field $\Psi_{fund}$ becomes real under the restricted transformation at $\tau =\beta$.  
In fact, $\Psi_{fund}$ and its charge conjugation $\Psi_{fund}^C$ 
are transformed under the restricted transformation as $\Psi_{fund}\to (U_{\rm Sf})^k\Psi_{fund}$ and $\Psi_{fund}^C\to (U_{\rm Sf})^k\Psi_{fund}^C$ at $\tau =\beta$. 
Obviously, this reality of $\Psi_{fund}$ at $\tau =\beta$ ensures $\mathbb{Z}_N$ symmetry to be exact. 

Now we consider the chiral limit of $m=0$ 
as an ideal case and take the case of $N=N_{F,fund} =3$ for a typical example. 
The present chiral- and $\mathbb{Z}_3$-symmetric $SU(3)$ gauge theory with FTBC fundamental fermions was constructed from 
the $SU(3)$ gauge theory in which fundamental fermions have 
$SU(3)_{\rm R}$ and $SU(3)_{\rm L}$ flavor symmetries. 
The $SU(3)_{\rm R}$ and $SU(3)_{\rm L}$ symmetries are broken down to $(U(1)_{\rm R})^2$ and $(U(1)_{\rm L})^2$ symmetries, respectively, by introducing the FTBC, i.e., 
the flavor-dependent imaginary chemical potential. 
These symmetries remain as continuous symmetries in addition to global $U(1)_{\rm B}$ symmetry.  
Thus, instead of the preservation of $\mathbb{Z}_3$ symmetry, the chiral symmetry is partially broken. 
This implies that there is a NO-GO theorem on construction of 
the chiral- and $\mathbb{Z}_N$-symmetric $SU(N)$ gauge theory with ordinary fundamental fermions. 
It seems that it is impossible to construct a fully chiral symmetric and $\mathbb{Z}_N$  symmetric $SU(N)$ gauge theory with fundamental fermions 
without introducing the additional gauge field 
that plays the role of the conjugate gauge field $\tilde{A}_\nu$ 
in the $SU(N)$ gauge theory with adjoint fermions and 
is coupled to other charges such as flavor or baryon number. 
Several properties of gauge theories with ordinary fundamental, FTBC fundamental and adjoint fermions are summarized 
in Table \ref{Table:property}.  

\begin{table}[h]
\begin{center}
\begin{tabular}{cccc}
\\
\hline
\hline
property~~ & ~~Fundmental~~ & ~~FTBC~~&~~Adjoint~~~\\
\hline
\hline
$\mathbb{Z}_N$ & broken & ~~symmetric~~&~~~symmetric \\
\hline
Reality &  not real       & not real   &real \\
\hline
Flavor & symmetric  & partially broken    &  symmetric \\
\hline
Chiral & symmetric  &  partially broken   & symmetric   \\
\hline
Sign problem &~~exist & ~~exist~~&~~~none for even $N_{F,adj}$ \\
\hline
\hline
\end{tabular}
\end{center}
\caption{ Summary of properties of $SU(N)$ gauge theories with 
ordinary fundamental, FTBC fundamental and adjoint fermions. 
Chiral symmetry is considered in the case of $m=0$. 
}
\label{Table:property}
\end{table}

Instead of the flavor-dependent imaginary chemical potential, 
one may consider the color-dependent imaginary chemical potential $i\theta^C T$ with  
\begin{eqnarray}
\theta^C&=&{\rm diag}(\theta^C_1,\theta^C_2,\cdots,\theta^C_N)
\nonumber\\
&=&{\rm diag}(0,2\pi/N,\cdots, 
\nonumber\\
&& 2(l-1)/\pi/N,\cdots,2\pi (N-1)). 
\label{CTBC}
\end{eqnarray}
In this case, the gauge symmetry is partially broken; 
for example, $SU(3)$ gauge symmetry is broken to $U(1)^2$ 
the generators of which are $t_3$ and $t_8$ in Cartan sub-algebra of $SU(3)$ group. 
However, the system is $\mathbb{Z}_3$ symmetric under the baryon-number-color linked (BCL) transformation
\begin{eqnarray}
\Psi_{fund} &\to& \Psi_{fund}^\prime =UV\Psi_{fund}
\nonumber\\
A_\nu &\to& A_\nu^\prime =UA_{\nu}U^{-1}+i(\partial_\nu U)U^{-1} , 
\label{ZNtrans_BC}
\end{eqnarray}
where 
\begin{eqnarray}
V(\tau )=\exp{(i2\pi k T\tau /3)}(U_{\rm Sc})^k 
\label{BCL-trans} 
\end{eqnarray}
and 
\begin{eqnarray}
&&U(x,\tau)=\exp[\{i(\alpha_3t_3+\alpha_8t_8)]
\label{color_3_8}
\end{eqnarray}
with the temporary boundary condition
\begin{eqnarray}
U (x,\beta )&=&\exp{(-i2\pi k/3)}U(x,0).
\\
\label{Z_N_twist_2}
\end{eqnarray}
The matrix $U_{\rm Sc}$ has the same form as $\hat{U}_{\rm Sf}$ but acts on color space. 
When $\tau$ is neither 0 nor $\beta$, 
$U$ is an element of color $SU(N)$ group, but $V$ is not. 
Therefore, this invariance is not trivial. 
Thus $\mathbb{Z}_3$ symmetry may appear by abandoning full $SU(3)$ gauge invariance.  

Even in the theory with no color-dependent  chemical potential, 
gauge symmetry is partially broken, 
if the temporal gauge field $A_4$ has an expectation value. 
In this situation, the expectation value acts just as the color-dependent imaginary chemical potential.  
At low energy where gauge interaction is strong, the GB mentioned above 
is expected to be canceled out by confinement~\cite{Kugo-Ojima}. 
At high energy where the gauge interaction is weak, 
in contrast, the breaking may appear. 
This is nothing but the Hosotani mechanism~\cite{Hosotani}.  
This possibility is discussed in Sec. \ref{sec:GB} particularly for 
the case of FTBC fundamental fermions.

\section{Low-energy dynamics}
\label{sec:PNJL}

In this section, we analyze low-energy dynamics of $SU(3)$ gauge theories with FTBC and adjoint (ADJ) fermions and 
investigate similarities between them, using the Polyakov-loop extended Nambu-Jona-Lasinio (PNJL) model~\cite{Meisinger,Dumitru,Fukushima,Ratti,Megias,Rossner,
Schaefer,Abuki,Fukushima2,Kashiwa1,McLerran_largeNc,Nishimura_ADJ,Kahara_ADJ,Zhang_ADJ,Sakai,Sasaki-T_Nf3}. 
The PNJL Lagrangian density with $N_F$ degenerate flavors is obtained by   
\bea
{\cal L}_{\rm PNJL}&=& \sum_f^{N_F}\bar{\Psi}_f(\gamma_\nu D^{\rm PNJL}_\nu +m)\Psi_f+{\cal L}_{\rm NJL}
+{\cal U},
\label{PNJL}
\eea
where $D^{\rm PNJL}_\nu=\partial_\nu -i\delta_{\nu,4}A_4$. 
We use the Polyakov gauge in which $A_i=0$ for $i=1,2,3$ and $A_4$ is considered as a background field.  

In (\ref{PNJL}), ${\cal L}_{\rm NJL}$ stands for effective quark-quark interactions. 
For fundamental quarks with $N=N_{F,fund}=3$, they have 
the standard form of 
\begin{eqnarray}
{\cal{L}}_{{\rm NJL},fund}
&=&-G_{\rm S}\sum_{a=0}^{8}[({\bar \Psi}\lambda_a \Psi)^2+({\bar \Psi}i\gamma_5\lambda_a \Psi )^2] 
\nonumber\\
&&+G_{\rm D}\left[\det_{ff^\prime}{\bar \Psi}_f(1+\gamma_5)\Psi_{f^\prime} +{\rm h.c.}\right], 
\label{NJL-F}
\end{eqnarray}
where $\lambda_a$ is the Gell-Mann matrix in flavor space and 
$G_{\rm S}$ and $G_{D}$ are coupling constants of the 
scalar-type four-quark interaction and the Kobayashi-Maskawa-t'Hooft determinant interaction, respectively~\cite{KMK,tHooft}. 
Table \ref{table-para_Nc3}(a) shows values of the coupling constants 
in addition to the current quark mass $m$ and the three-dimensional momentum cutoff $\Lambda$.  
The coupling constants and the cutoff are determined to reproduce 
empirical values of $\eta'$- and $\pi$-meson masses and $\pi$-meson decay constant at vacuum when $m_u=m_d=5.5$MeV and $m_s=140.7$MeV~\cite{Rehberg}. 
In this paper, however, we take symmetric current quark masses,  $m_f=5.5$MeV for any flavor, to make the system flavor symmetric 
at $\theta_f=0$. 

\begin{table}[h]
\begin{center}
\begin{tabular}{c|ccccc}
\hline \hline
\raisebox{-1.5ex}[0cm][0cm]{(a)}&~$m_f$(MeV)~&~$\Lambda$(MeV)~&~$~G_{\rm s}\Lambda^2~$~&~$~G_{\rm D}\Lambda^5~$~&
\\
\cline{2-6}
&5.5&602.3&1.835&12.36&
\\
\hline \hline
\raisebox{-1.5ex}[0cm][0cm]{(b)}&~$m$(MeV)~&~$\Lambda$(MeV)~&~$~G\Lambda^2~$~&
\\
\cline{2-6}
&5.5&651& 3.607 &
\\
\hline \hline
\raisebox{-1.5ex}[0cm][0cm]{~(c)~}&$a_0$&$a_1$&$a_2$&$b_3$&$T_0$(MeV)
\\
\cline{2-6}
& 3.51 & -2.47 & 15.2 & -1.75 & 195
\\
\hline \hline
\end{tabular}
\caption{
Summary of the parameter set in the PNJL model: 
(a) parameters of the NJL sector for fundamental fermions with $N=3$ and $N_{F,fund}=3$, 
(b) parameters of the NJL sector for adjoint fermions with $N=3$ and $N_{F,adj}=1$, and  
(c) parameters of the Polyakov-loop potential for the case of $N=3$. }
\label{table-para_Nc3}
\end{center}
\end{table}

As the NJL sector for ADJ fermion,  we take  the form of 
\begin{eqnarray}
{{\cal L}}_{{\rm NJL},adj}=-G(\bar{Q}Q)^2
\label{NJL}
\end{eqnarray}
proposed in Ref.~\cite{Zhang_ADJ}; see 
Table \ref{table-para_Nc3}(b) for the parameter set in the case of $N=3$ and $N_{F,adj}=1$. 

In (\ref{PNJL}), the Polyakov-loop potential ${\cal U}$ 
is a function of $A_4$.  Particularly in the cases of $N=2$ and 3, it can be written as a function of the Polyakov-loop $\Phi$ and its conjugate $\Phi^*$~\cite{Polyakov}.   
In the fundamental representation, they are defined by
\begin{align}
\Phi &= {1\over{N}}{\rm tr}_c(L_{fund}),\quad
\Phi^* ={1\over{N}}{\rm tr}_c({\bar L_{fund}}),
\label{Polyakov_ncN}
\end{align}
where $L_{fund}$ is the fundamental representation of $L=\exp(i\phi)= \exp(i A_4/T)$. 
In the Polyakov gauge, they are written as 
\begin{eqnarray}
\Phi ={1\over{N}}(e^{i\phi_1}+e^{i\phi_2}+\cdots +e^{i\phi_N}), 
\end{eqnarray}
where the $\phi_i$ satisfy the condition $\phi_1+\phi_2+\cdots +\phi_N=0$.  

For $N=3$, we take the Polyakov-loop potential ${\cal U}$ of Ref.~\cite{Rossner}:
\begin{align}
&{\cal U} = T^4 \Bigl[-\frac{a(T)}{2} {\Phi}^*\Phi\notag\\
      &~~~~~+ b(T)\ln(1 - 6{\Phi\Phi^*}  + 4(\Phi^3+{\Phi^*}^3)
            - 3(\Phi\Phi^*)^2 )\Bigr] ,
            \label{eq:E13}\\
&a(T)   = a_0 + a_1\Bigl(\frac{T_0}{T}\Bigr)
                 + a_2\Bigl(\frac{T_0}{T}\Bigr)^2,~~~~
b(T)=b_3\Bigl(\frac{T_0}{T}\Bigr)^3 .
            \label{eq:E14}
\end{align}
Parameters of $\mathcal{U}$ are fitted to LQCD data 
at finite $T$ in the pure gauge limit. 
The Polyakov-loop potential yields the first-order deconfinement phase transition 
at $T=T_0$ in the pure gauge theory~\cite{Boyd,Kaczmarek}. 
The original value of $T_0$ is $270$ MeV determined from the pure gauge 
LQCD data, but the PNJL model with this value yields a larger 
value of the pseudocritical temperature $T_\mathrm{c}$ 
at zero chemical potential than $T_c\approx 160$~MeV predicted 
by full LQCD \cite{Borsanyi,Soeldner,Kanaya}. 
We then rescale $T_0$ to 195~MeV so as to reproduce 
$T_c\sim 160$~MeV~\cite{Sasaki-T_Nf3}. 
Parameters in the Polyakov-loop potential are summarized in Table~\ref{table-para_Nc3}(c), together with the other parameters. 

For the case of $N_{F,fund}=N_NN~(N_N=1,2,\cdots )$ fundamental quarks with the FTBC,  
the thermodynamic potential $\Omega_{\rm FTBC}$ is obtainable 
from ${\cal L}_{\rm PNJL}$ with the mean-field approximation: 
\begin{align}
\Omega_{\rm FTBC}
&=N_N\Omega_{q,\rm FTBC}+U_{fund}+{\cal U}
\label{PNJL_Omega_TBC}
\end{align}
with 
\begin{align}
\Omega_{q,\rm FTBC}&= -2\sum_{c=1}^N \sum_{f  =1}^N\int \frac{d^3 p}{(2\pi)^3}
\Bigl[E_f \notag\\
&+ \frac{1}{\beta}\ln~ [1 + e^{i\phi_c}e^{i\theta_{f} }e^{-\beta E_{f-}}]
\notag\\
&    + \frac{1}{\beta}\ln~ [1 + e^{-i\phi_c}e^{-i\theta_{f} }e^{-\beta E_{f+}}]
\Bigr]
\notag\\
&+U_{fund}+{\cal U}, 
\label{PNJL-Omega_q_TBC}
\end{align}
where $E_f=\sqrt{{\bf p}^2+M_f^2}$ 
and $E_{f\pm}=\sqrt{{\bf p}^2+M_f^2}\pm i(\pi  -\varphi )T$ for 
the constituent quark mass $M_f$. 
For $N=N_{F,fund}=3$, the constituent quark mass $M_f$ and the mesonic potential $U_{fund}$ are given by
\begin{eqnarray}
M_f=m_f-4G_{\rm S}\sigma_f+2G_{\rm D}\sigma_{f^\prime}\sigma_{f^{\prime\prime}}
\label{Mass}
\end{eqnarray}
for $f\neq f^\prime$ and $f\neq f^{\prime\prime}$ and $f^\prime =f^{\prime\prime}$
and 
\begin{eqnarray}
U_{fund}^{N=N_{F,fund}=3}=\sum_{f=u,d,s}2G_{\rm S}\sigma_f^2-4G_{\rm D}\sigma_u\sigma_d\sigma_s,
\label{MP}
\end{eqnarray}
where $\sigma_f~(f=u,d,s)$ is the expectation value of chiral condensate $\bar{\Psi}_f\Psi_f$. 

For $N=3$, when the summation is taken over color indices $c$, 
Eq. (\ref{PNJL-Omega_q_TBC}) is rewritten into 
\begin{align}
\Omega_{q,\rm FTBC}^{N=3}
&= -2 \sum_{f  =1}^3\int \frac{d^3 p}{(2\pi)^3}
\Bigl[3E_f \notag\\
&+ \frac{1}{\beta}\ln~ [1 + 3\Phi e^{i\theta_{f} }e^{-\beta E_{f-}}
\nonumber\\
&
+3\Phi^* e^{2i\theta_{f} }e^{-2\beta E_{f-}}+e^{3i\theta_{f}}e^{-3\beta E_{f-}}]
\notag\\
&    + \frac{1}{\beta}\ln~ [1 + 3\Phi^* e^{-i\theta_{f} }e^{-\beta E_{f+}}
\nonumber\\
&+ 3\Phi e^{-2i\theta_{f} }e^{-2\beta E_{f+}}+e^{-3i\theta_{f}}e^{-3\beta E_{f+}}].
\Bigr]
\label{PNJL-Omega_TBC_2}
\end{align}
If $\Phi =\Phi^*=0$, this equation is further reduced to
\begin{align}
\Omega_{q,\rm FTBC}^{N=3}
&= -2 \sum_{f  =1}^3\int \frac{d^3 p}{(2\pi)^3}
\Bigl[3E_f \notag\\
&+ \frac{1}{\beta}\ln~ [1 +e^{-3\beta E_{f-}}]
\notag\\
&    + \frac{1}{\beta}\ln~ [1 +e^{-3\beta E_{f+}}]
\Bigr].
\label{PNJL-Omega_TBC_3}
\end{align}
Equation \eqref{PNJL-Omega_TBC_3} shows that red, blue and green quarks are statistically in the same state. 
The flavor-dependent imaginary chemical potential disappears 
in  \eqref{PNJL-Omega_TBC_3} for the confined phase of $\Phi =\Phi^*=0$.  
The flavor-symmetry breaking due to the flavor-dependent imaginary chemical potential is thus dynamically restored in the confined phase.  
As a consequence of this property, Eq. \eqref{PNJL-Omega_TBC_3} has a simpler form of 
\begin{align}
\Omega_{q,\rm FTBC}^{N=3}
&= -2 \int \frac{d^3 p}{(2\pi)^3}
\Bigl[9E_{\rm con} \notag\\
&+ \frac{3}{\beta}\ln~ [1 +e^{-i3(\pi -\varphi)} e^{-3\beta E_{\rm con}}]
\notag\\
&    + \frac{3}{\beta}\ln~ [1 +e^{i3(\pi-\varphi)}e^{-3\beta E_{\rm con}}]
\Bigr], 
\label{PNJL-Omega_TBC_5}
\end{align}
with $M_{\rm con}\equiv M_u=M_d=M_s$ and $E\equiv \sqrt{{\bf p}^2+M_{\rm con}^2}$ in confined phase.

Now we consider ADJ fermions with 
$N_{F,adj}=N_{F,fund}/N = N_N$ flavors. 
For the fermion boundary condition with $\varphi$,   
the thermodynamic potential is obtained 
with the mean field approximation as 
\begin{align}
\Omega_{adj}
&=N_N\Omega_{q,adj}+U_{adj}+{\cal U} 
\label{PNJL-Omega_adj_total}
\end{align}
with  
\begin{align}
\Omega_{q,adj}
&= -2\sum_{c=1}^N \sum_{c^\prime  =1}^N\int \frac{d^3 p}{(2\pi)^3}
\Bigl[E \notag\\
&+ \frac{1}{\beta}\ln~ [1 + e^{i\phi_c}e^{-i\phi_{c^\prime} }e^{-\beta E_-}]
\notag\\
&    + \frac{1}{\beta}\ln~ [1 + e^{-i\phi_c}e^{i\phi_{c^\prime} }e^{-\beta E_+}]
\Bigr]
\notag\\
&-\Omega_{1},
\label{PNJL-Omega_adj}\\
\Omega_1 &=-2 \int \frac{d^3 p}{(2\pi)^3}
   \Bigl[ E 
\notag\\
&+\sum_{j=\pm}\frac{1}{\beta}\ln~ [1 + e^{-\beta E_j}]\Bigr] \Bigr] 
\label{omega_1} ,
\end{align}
$M=m-2G_{\rm S}\sigma$, $E=\sqrt{{\bf p}^2+M^2}$, $E_\pm =E\pm i(\pi -\varphi)T$ and $U_{adj}=G \sigma^2$ for 
the expectation value $\sigma$ of chiral condensate $\bar{\Psi}\Psi$.   

For $N=3$, when the summation is taken over color indices $c$ and $c^\prime$, Eq. (\ref{PNJL-Omega_adj}) becomes~\cite{Kahara_ADJ}
\begin{align}
\Omega_{q,{adj}}^{N=3}
&= -2 \int \frac{d^3 p}{(2\pi)^3}
   \Bigl[ 8E
\notag\\
&         + \sum_{j=\pm}\frac{1}{\beta}\ln~ [1 + (9\Phi\Phi^*-1)e^{-\beta E_j}
\notag\\
&+(27\Phi^3+27{\Phi^*}^3-27\Phi\Phi^*+1)e^{-2\beta E_j}
\notag\\
&+(81\Phi^2{\Phi^*}^2-27\Phi\Phi^*+2)e^{-3\beta E_j}
\notag\\
&+(162\Phi^2{\Phi^*}^2-54\Phi^3-54{\Phi^*}^3+18\Phi\Phi^*-2)e^{-4\beta E_j}
\notag\\
&         +  (81\Phi^2{\Phi^*}^2-27\Phi\Phi^*+2)e^{-5\beta E_j}
\notag\\
&+(27\Phi^3+27{\Phi^*}^3-27\Phi\Phi^*+1)e^{-6\beta E_j}
\notag\\
& +(9\Phi\Phi^*-1)e^{-7\beta E_j}  +e^{-8\beta E_j}
]
\Bigr]. 
\label{PNJL-adj_nc3}
\end{align}
The one-quark state, i.e., the term proportional to $e^{-\beta E_j}$ 
does not vanish in (\ref{PNJL-adj_nc3}), even if $\Phi =\Phi^*=0$. 
If $\Phi =\Phi^*=\pm 1/3$, meanwhile, the one-quark state vanishes and the Polyakov-loop $\Phi_{adj}$ in the adjoint representation, given by~\cite{Kahara_ADJ} 
\begin{eqnarray}
\Phi_{adj}={1\over{8}}{\rm tr}(L_{adj})={1\over{8}}(9\Phi\Phi^*-1),
\label{eqnarray}
\end{eqnarray}
also vanishes, where $L_{adj}$ are the adjoint representation of $L$.  
As seen below, however, this situation is not realized. 
Besides, even if it is realized, one can not consider a confinement with 
$\Phi_{adj}$, since $\Phi_{adj}$ 
is invariant under $\mathbb{Z}_3$ transformation and hence 
it is not an order parameter of $\mathbb{Z}_3$ symmetry.  
We then consider a confinement as that of the static fundamental 
charge. 
For this definition of confinement, we can use $\Phi$ as an order parameter of a confinement/deconfinement transition. 
In fact, LQCD simulations~\cite{KL_aQCD} 
of a $SU(3)$ gauge theory with ADJ fermions indicate 
that below the critical temperature of the deconfinement transition   
a potential between static fundamental charges 
linearly raises with respect to increasing a distance $R$ between the 
two charges, but a potential between static adjoint charges has a 
linearly-raising form only at small $R$. Thus a string breaking 
occurs at large $R$ when static adjoint charges are taken. 

\begin{figure}[htbp]
\begin{center}
\includegraphics[width=0.3\textwidth]{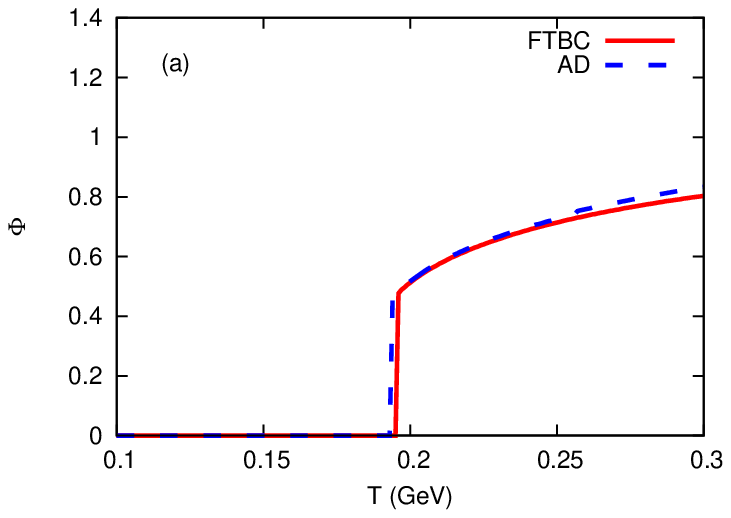}
\includegraphics[width=0.3\textwidth]{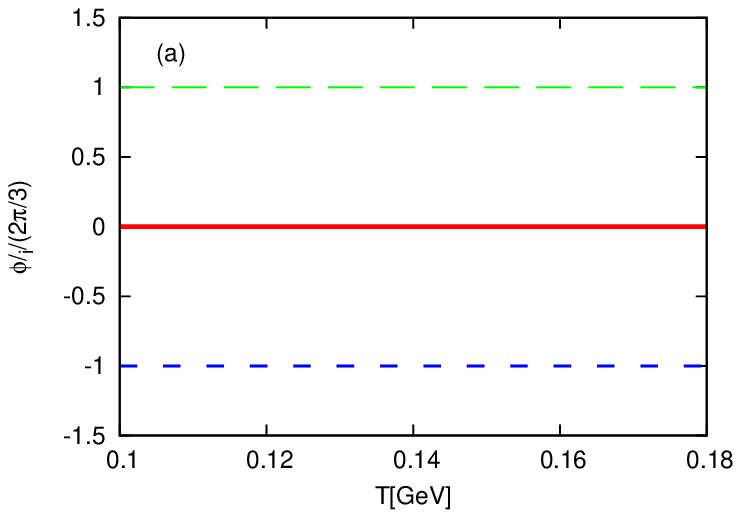}
\includegraphics[width=0.3\textwidth]{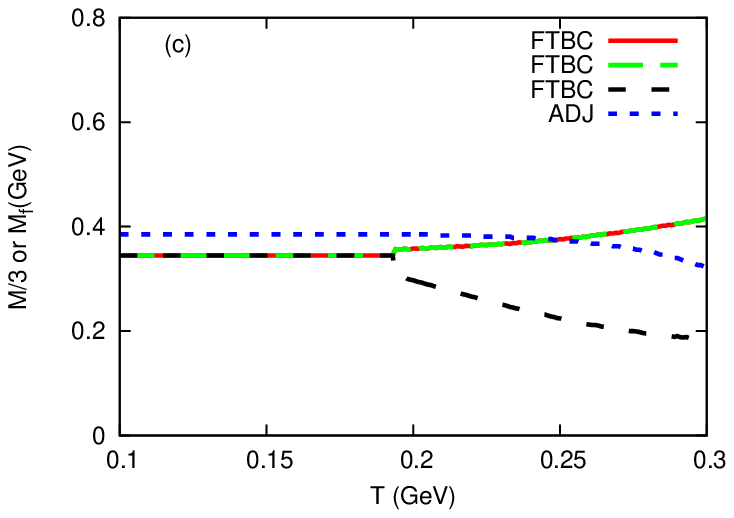}
\end{center}
\caption{$T$ dependence of order parameters 
in the PNJL model of 
FTBC fermion with $N=N_{F,fund}=3$ and that of ADJ fermion 
with $N=3$ and $N_{F,adj}=1$. 
In the both cases, the boundary condition $\varphi =\pi$ is taken.
Three panels correspond to (a) $\Phi$, (b) $\phi_c$, and (c) $M/3$ and $M_f$, respectively.  
}
\label{PNJL_order}
\end{figure}

In numerical calculations of the PNJL model, we take $N=3$ and $N_3=1$, i.e., $N_{F,fund}=3$ and $N_{F,adj}=1$. 
The expectation values $\sigma$ (or $\sigma_f$) and the $\phi_c$ are determined by the minimum condition of $\Omega$. 
Figure~\ref{PNJL_order}(a) shows $T$ dependence of $\Phi$ 
for FTBC fermion with $N=N_{F,fund}=3$ and ADJ fermion 
with $N=3$ and $N_{F,adj}=1$. 
Here we take the anti-periodic boundary condition by setting 
$\varphi =\pi$. 
The two cases show similar $T$ dependence: 
more precisely, $\Phi$ has a jump at $T=T_c\sim 195$MeV from 0 to 0.5.
Below $T_c$, $\mathbb{Z}_3$ symmetry is surely preserved 
in both the cases. 
For the both cases, $\Phi$ never has $\pm 1/3$, since 
$\Phi$ has a jump from 0 to 0.5. 

Figure~\ref{PNJL_order}(b) shows $T$ dependence of the $\phi_c~(c=1,2,3)$ at low $T$ for the two cases after an appropriate relabeling of color indexes $c$. It is found 
that $\phi_c=2\pi k/3$ with $k=0,\pm 1$ at low $T$. 
The results for ADJ and FTBC fermions agree with each other. 
Therefore, 
\begin{eqnarray}
\phi_c^{adj}=\phi_c^{fund}=-\theta_f
\label{condition}
\end{eqnarray}
holds true after an appropriate relabeling of $c$. 
This means that the back ground gauge field in (\ref{PNJL-Omega_q_TBC}) is the same as in  (\ref{PNJL-Omega_adj}). 
The two models thus have similar dynamics to each other at low $T$, 
as far as 
the confinement/deconfinement transition is concerned. 

On the other hand, the chiral property is somewhat different between the two models. 
Figure~\ref{PNJL_order}(c) shows $T$ dependence of $M$ and $M_f$ for the two cases. 
The constituent quark mass $M$ of ADJ fermion is much larger 
than $M_f$ of FTBC fermion.  
For FTBC fermion, furthermore, three degenerate quarks split into two heavy ones and light one at high $T$, since flavor symmetry is broken 
by the $\mathbb{Z}_3$-symmetry breaking. 
It is found from discussions in Sec. \ref{fermion} 
that the differences in the chiral sector are originated in the presence or absence of fluctuations of the conjugate gauge field $\tilde{A}_\nu$ around its means value. 

As mentioned above, the constituent quark mass $M$ of adjoint fermion is much heavier than $M_f$ of FTBC fermion at low $T$. 
However, there is a approximate scaling low $M/M_f\sim N_c$.  
This can be understood as follows. 
From the stationary conditions for $\sigma$ and $\sigma_f$ at $T=0$, 
we can obtain the following equation, 
\begin{eqnarray}
\sigma =-{N_{adj}\over{\pi^2}}\int_0^{\Lambda_{adj}}dpp^2{M\over{E}}\sim -{N_{adj}\over{3\pi^2}}\Lambda_{adj}^3, 
\label{sigma_adj}\\
\sigma_f=-{N\over{\pi^2}}\int_0^{\Lambda_{fund}}dpp^2{M_f\over{E_f}}\sim -{N\over{3\pi^2}}\Lambda_{fund}^3, 
\label{sigma_FTBC}
\end{eqnarray}
where $N_{adj}$ is the dimension of the adjoint representation and 
is related to $N$ as $N_{adj}=N^2-1$. 
Since $\Lambda_{adj}\sim \Lambda_{fund}$, we obtain
$M/M_f\sim \sigma/\sigma_f\sim N_{adj}/N\sim N=3$. 
Therefore, $M$ is approximately three times heavier than $M_f$. 
For FTBC fermion, only the three-quark state can survive 
in the confinement phase, as shown in (\ref{PNJL-Omega_TBC_5}).  
Therefore, the quark part of the thermodynamic potential 
has a small contribution and do not affect much the confinement/deconfinement transition in the two cases. This is the reason why 
the confinement/deconfinement transitions are similar to each other 
between ADJ and FTBC matters.

\begin{figure}[htbp]
\begin{center}
\includegraphics[width=0.3\textwidth]{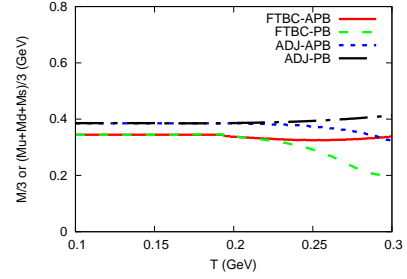}
\end{center}
\caption{$T$ dependence of constituent quark masses in 
different types of fermions and boundary conditions. 
Four cases of FTBC-APB, FTBC-PB, ADJ-APB and ADJ-PB are 
taken, where FTBC and ADJ stand for kinds of fermions while 
PB and APB correspond to kinds of boundary conditions. 
Here we set $N=N_{F,fund}=3$ and $N_{F,adj}=1$.  
For FTBC-APB and FTBC-PB, 
constituent quark masses depend on flavor 
in the deconfinement phase at higher temperature, 
so the average vales are shown in the cases. 
} 
\label{PNJL_M_APB_PB}
\end{figure}

Putting $\Phi =0$ and neglecting higher-order terms of the suppression factor $e^{-\beta E}$,  we obtain 
\begin{align}
& \Omega_{q,{adj}}^{N=3}
= -2 \int \frac{d^3 p}{(2\pi)^3}
   \Bigl[ 8E
\notag\\
& + \frac{1}{\beta}[\ln~ [1 -e^{-i(\pi -\varphi)}e^{-\beta E}]
+\frac{1}{\beta}\ln~ [1 -e^{i(\pi -\varphi)}e^{-\beta E}]\Bigr]. 
\label{PNJL-adj_nc3_2}
\end{align}
Comparing \eqref{PNJL-adj_nc3_2} with (\ref{PNJL-Omega_TBC_5}) and noting $M \sim 3M_f$ in the confined phase, we see that the thermodynamic potential of ADJ fermion with $\varphi=0~(\pi)$ mimics that of FTBC fermion with $\varphi =\pi (0)$ in the confined phase, although coefficients of terms in 
\eqref{PNJL-adj_nc3_2} are somewhat different 
from those in (\ref{PNJL-Omega_TBC_5}). 
Since the fermion contribution itself is small 
in the confined phase, the difference between 
the periodic boundary condition (PB) and the anti-periodic boundary condition (APB) is negligibly small 
at low $T$, as shown in Fig.~\ref{PNJL_M_APB_PB}. 

\section{Gague symmetry breaking in FTBC model}
\label{sec:GB}

As far as the confinement/deconfinement transition is concerned, there exist a similarity between FTBC and ADJ fermions even at high $T$, 
as shown in \ref{PNJL_order}(a). 
This implies the possibility that the  gauge symmetry breaking (GB) at higher temperature or higher-energy scale takes place also for FTBC fermion as a result of the Hosotani mechanism~\cite{Hosotani,Kashiwa_GB}. 
In this section, we examine the GB in $SU(3)$ gauge theories with 
different types of fermions and boundary conditions 
on $R^3\times S^1$ by using the one-loop effective potential.  
For simplicity, we regard the compact Euclidean time direction $\tau$ as a spacial direction $y$ and replace $1/T$ by a size $L$ of $S^1$. 
The gauge field is then obtained as 
\begin{eqnarray}
A_\mu = \langle A_y \rangle +A^\prime_\mu, 
\label{Ay}
\end{eqnarray}
where the VEV of $A_y$, $\langle A_y \rangle$, can be written as
\begin{eqnarray}
\langle A_y \rangle={2\pi\over{L}}q
\label{q_color}
\end{eqnarray}
where $q={\rm diag}(q_1,q_2,q_3)$ with $q_1+q_2+q_3=0$ and each component is determined as $(q_i)_{{\rm mod}~1}$.  
As in the previous section, the Polyakov-loop $\Phi$ is defined as
\begin{eqnarray}
\Phi ={1\over{3}}(e^{i2\pi q_1}+e^{i2\pi q_2}+e^{i2\pi q_3})
\label{Polyakov-loop}
\end{eqnarray}
Here we call the $q_i$ "Polyakov-loop phases".  
When the solution is nontrivial, i.e., when $q$ does not have a form 
$at_0$ for a constant $a$ and the unit matrix $t_0$ in color space, 
the GB is expected to take place.  

The one-loop effective potential $\mathcal{V}$ is a function of the $q_i$ 
and consists of the gluon part $\mathcal{V}_{g}$ and the fermion part $\mathcal{V}_{f}$:
\begin{eqnarray}
\mathcal{V}&=\mathcal{V}_{g}+\mathcal{V}_{f}.  
\label{effective-V}
\end{eqnarray}
The gluon part $\mathcal{V}_{g}$ has an explicit form of   
\begin{eqnarray}
{\cal V}_{g}
=-{2\over{L^4\pi^2}}\sum_{i=1}^3\sum_{j=1}^3\sum_{n=1}^{\infty}
\Bigl( 1 - \frac{1}{3} \delta_{ij} \Bigr)
\frac{\cos( 2 n \pi q_{ij})}{n^4} 
\label{V_gauge}
\end{eqnarray}
with $q_{ij} = ( q_i - q_j )_{{\rm mod}~1}$.  
 For the case of fundamental fermion,  ${\cal V}_{f}$ has a form of 
\begin{eqnarray}
{\cal V}_{f,fund}
&=& \frac{2N_{F,fund}m^{2}}{L^2 \pi^2} \sum_{i=1}^3 \sum_{n=1}^\infty
{K_{2}(nmL)\over{n^{2}}} 
\nonumber\\
&&\times \cos[2 \pi n (q_i+\varphi )]  
\label{V_FD}   
\end{eqnarray}
for the fermion mass $m$ and  the boundary angle $\varphi $ defined in 
\eqref{BC}, where $K_2$ is the modified Bessel function 
of the second kind. 
 In the case of fundamental FTBC fermion, 
${\cal V}_{f}$ is modified as 
\begin{eqnarray}
{\cal V}_{f,\rm FTBC}
&=& \frac{2N_{3}m^{2}}{L^2 \pi^2} \sum_{i=1}^3 \sum_{f=1}^3\sum_{n=1}^\infty
\nonumber\\
&&\times {K_{2}(nmL)\over{n^{2}}}\cos[2 \pi n (Q_{if}+\varphi )],  
\label{V_FTBC}   
\end{eqnarray}
where $N_3=N_{F,fund}/3$ and $Q_{if}=q_i+(f-1)/3$.   
 In the case of ADJ fermion,  meanwhile, ${\cal V}_{f}$ has a form of 
\begin{eqnarray}
{\cal V}_{f,adj}
&=& \frac{2N_{F,adj}m^{2}}{L^2 \pi^2} \sum_{i=1}^3 \sum_{j=1}^3\sum_{n=1}^\infty
\Bigl( 1 - \frac{1}{3} \delta_{ij} \Bigr)
\nonumber\\
&&\times {K_{2}(nmL)\over{n^{2}}} \cos[2 \pi n (q_{ij}+\varphi )].   
\label{V_ADJ}   
\end{eqnarray}

Now we consider six combinations of different fermions and boundary conditions. 

\begin{enumerate}
\item "FD-APB": fundamental fermions with anti-periodic boundary condition $\varphi=\pi$
\item  "FD-PB": fundamental fermions with periodic boundary condition $\varphi=0$ 
\item "ADJ-APB": adjoint fermions with anti-periodic boundary condition $\varphi=\pi$
\item  "ADJ-PB": adjoint fermions with periodic boundary condition $\varphi=0$
\item "FTBC-APB": FTBC fundamental fermions with 
the boundary condition $\varphi=\pi$ 
\item "FTBC-PB":  FTBC fundamental fermions with 
the boundary condition $\varphi=0$
\end{enumerate}

\begin{figure}[htbp]
\begin{center}
\includegraphics[width=0.3\textwidth]{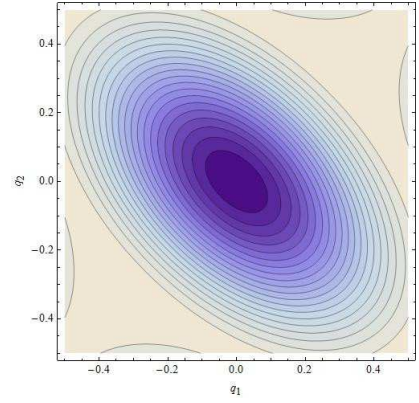}
\end{center}
\caption{Contour plot of ${\cal V}_{f}L^4$ in the limit $mL\to 0$ 
for the case of FD-APB.  Here, $q_3$ is given by $-q_1-q_2$.   
}
\label{FD_APB}
\end{figure}

\begin{figure}[htbp]
\begin{center}
\includegraphics[width=0.3\textwidth]{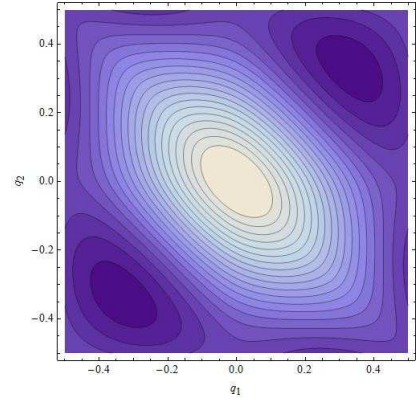}
\end{center}
\caption{The same figure as Fig. \ref{FD_APB} but for 
the case of FD-PB.    
}
\label{FD_PB}
\end{figure}

First, we investigate the structure of ${\cal V}_{f}$ in the high-energy limit $mL\to 0$.  
Figure~\ref{FD_APB} shows the contour plot of ${\cal V}_{f}$ 
for the case of FD-APB. 
The  fermion one-loop potential ${\cal V}_{f}$ becomes minimum at $(q_1,q_2,q_3)=(0,0,0)$ that 
gives $\Phi =1$ and hence can not induce the GB.  
For the case of FD-PB shown in Fig.\ref{FD_PB}, 
${\cal V}_{f}$ has minima at $(q_1,q_2,q_3)=(\pm 1/3,\pm 1/3,\mp 2/3)$ 
that lead to $|\Phi  |=1$; here  $(\pm 1/3,\pm 1/3,\mp 2/3)$ is a shorthand notation of   $(1/3,1/3,-2/3)$ and  $(-1/3,-1/3,2/3)$. 
Since $(\pm 1/3,\pm 1/3,\mp 2/3)=(\pm 1/3,\pm 1/3,\pm 1/3)_{{\rm mod}~1}$, the GB can not take place also in this case. 

\begin{figure}[htbp]
\begin{center}
\includegraphics[width=0.3\textwidth]{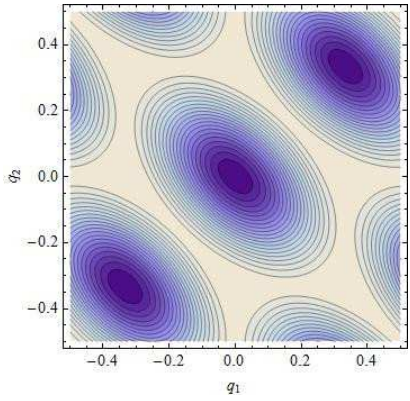}
\end{center}
\caption{The same figure as Fig. \ref{FD_APB} but for 
the case of ADJ-APB.    
}
\label{ADJ_APB}
\end{figure}

For the case of ADJ-APB, as shown in Fig.~\ref{ADJ_APB},
${\cal V}_{f}$ becomes minimum at $(q_1,q_2,q_3)=(0,0,0)$, $(\pm 1/3,\pm 1/3,\mp 2/3)$ that gives $|\Phi |=1$ and can not induce the GB. 
The fact that the three solutions are degenerate means that $\mathbb{Z}_3$ symmetry is preserved.     

\begin{figure}[htbp]
\begin{center}
\includegraphics[width=0.3\textwidth]{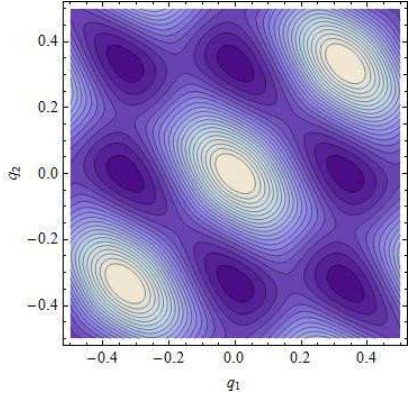}
\end{center}
\caption{The same figure as Fig. \ref{FD_APB}
but for the case of ADJ-PB.}
\label{ADJ_PB}
\end{figure}

For the case of ADJ-PB, as shown in Fig.~\ref{ADJ_PB}, 
${\cal V}_{f}$ has minima at $(q_1,q_2,q_3)=(\pm 1/3,\mp 1/3,0)$, $(\pm 1/3,0,\mp 1/3)$ and $(0,\pm 1/3,\mp 1/3)$. 
These solutions are a family in the sense that 
$(\pm 1/3,0,\mp 1/3)$ and $(0,\pm 1/3,\mp 1/3)$ are $\mathbb{Z}_3$ images of $(\pm 1/3,\mp 1/3,0)$. Since the family 
yields $\Phi =0$, this phase is sometimes called "re-confined phase".  
In the phase, the GB takes place and $SU(3)$ gauge symmetry is broken 
down to $U(1)\times U(1)$. 

\begin{figure}[htbp]
\begin{center}
\includegraphics[width=0.3\textwidth]{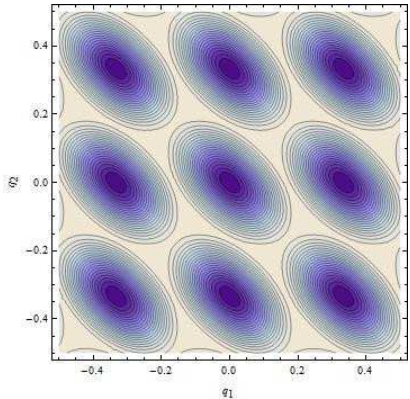}
\end{center}
\caption{
The same figure as Fig. \ref{FD_APB}
but for the case of FTBC-APB. 
}
\label{FTBC_APB}
\end{figure}

For the case of FTBC-APB, as shown in Fig.~\ref{FTBC_APB}, 
${\cal V}_{f}$ has minima at six points. 
The first $\mathbb{Z}_3$ family of $(q_1,q_2,q_3)=(0,0,0)$ and $(\pm 1/3,\pm 1/3,\mp 2/3)$ leads to $|\Phi  |=1$, whereas the second $\mathbb{Z}_3$ family of $(q_1,q_2,q_3)=(\pm 1/3,\mp 1/3,0)$, $(\pm 1/3,0,\mp 1/3)$ and 
$(0,\pm 1/3,\mp 1/3)$ yields $\Phi =0$. 
Thus $\mathbb{Z}_3$ symmetry is realized in FTBC fermions.  
The first family does not induce the GB, but 
the second family does. 

One can see from Figs.~\ref{ADJ_PB} and \ref{FTBC_APB} that 
near points yielding $\Phi=0$ the structure of ${\cal V}_f$ is 
similar between ADJ-PB and FTBC-APB fermions. 
This result is consistent with that in the previous section. 
Near points yielding $\Phi=1$, meanwhile, the structure of ${\cal V}_f$ is different between FTBC-APB and  ADJ-PB fermions.
The difference is important for the presence or absence of the GB in the total system, as seen later.  

\begin{figure}[htbp]
\begin{center}
\includegraphics[width=0.3\textwidth]{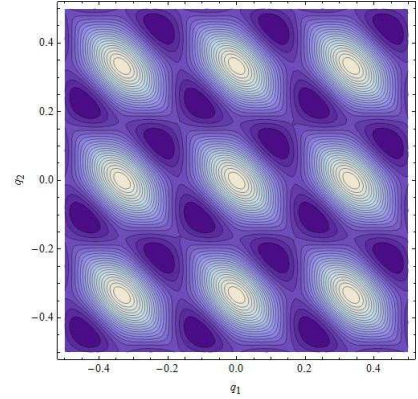}
\end{center}
\caption{
The same figure as Fig. \ref{FD_APB}
but for the case of FTBC-PB. 
}
\label{FTBC_PB}
\end{figure}

For the case of FTBC-PB, as shown in Fig.~\ref{FTBC_PB}, 
${\cal V}_{f}$ becomes minimum at $(q_1,q_2,q_3)=(\pm 1/9,\pm 1/9,\mp 2/9)$, $(\pm 2/9,\pm 2/9,\mp 4/3)$, $(\pm 4/9,\pm 4/3,\mp 8/9)$. The $\mathbb{Z}_3$ family of solutions yields a common value $|\Phi |=0.577$, 
that is, the GB takes place there. 
In the GB phase, $SU(3)$ gauge symmetry is broken to $SU(2)\times U(1)$ and $\mathbb{Z}_3$ and charge-conjugation symmetry are spontaneously broken. 

\begin{figure}[htbp]
\begin{center}
\includegraphics[width=0.3\textwidth]{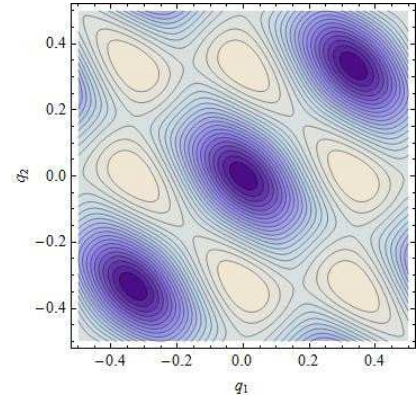}
\end{center}
\caption{Contour plot of ${\cal V}_{g}L^4$ in the  $q_1$-$q_2$ plane.   
Here, $q_3$ is given by $-q_1-q_2$.  
}
\label{G_boson}
\end{figure}

Next we consider the gluon one-loop potential  ${\cal V}_{g}$. 
As shown in Fig.~\ref{G_boson}, the  potential ${\cal V}_{g}$ has minima 
at three solutions $(q_1,q_2,q_3)=(0,0,0),(\pm 1/3,\pm 1/3,\mp 2/3)$ leading to $|\Phi  |=1$. 
In the case of FTBC-APB, therefore, the GB does not take place when 
${\cal V}_{g}$ is switched on, since ${\cal V}_{g}$ makes the solutions 
yielding $|\Phi |=1$ more stable than the solutions yielding $\Phi =0$.  
In the case of ADJ-PB, the re-confined phase appears even for $N_{F,adj}=1$, since effects of ${\cal V}_{f}$ overcome those of ${\cal V}_{g}$ 
in the small $mL$ limit, .   

\begin{figure}[htbp]
\begin{center}
 \includegraphics[width=0.3\textwidth]{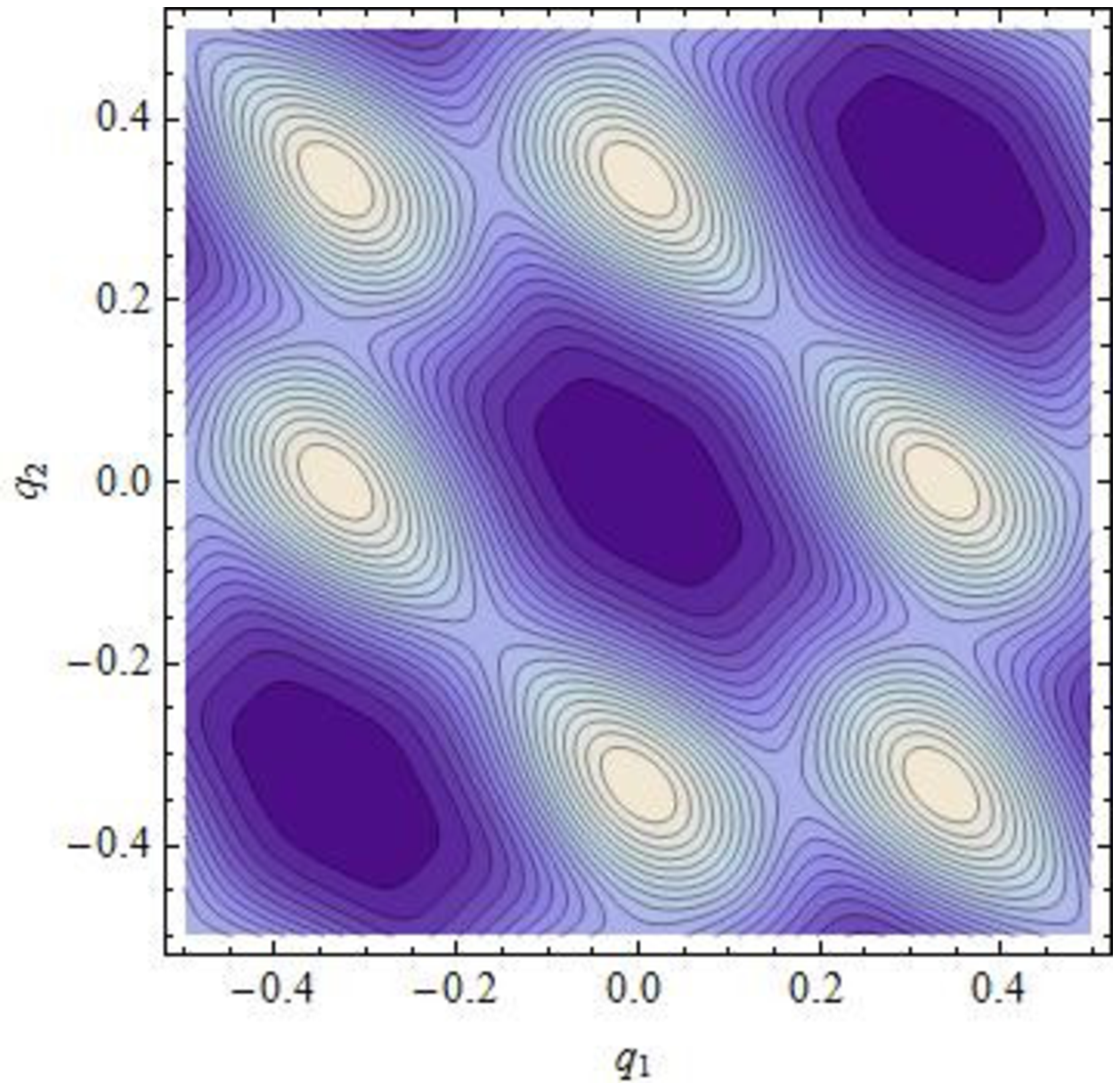}
 \includegraphics[width=0.3\textwidth]{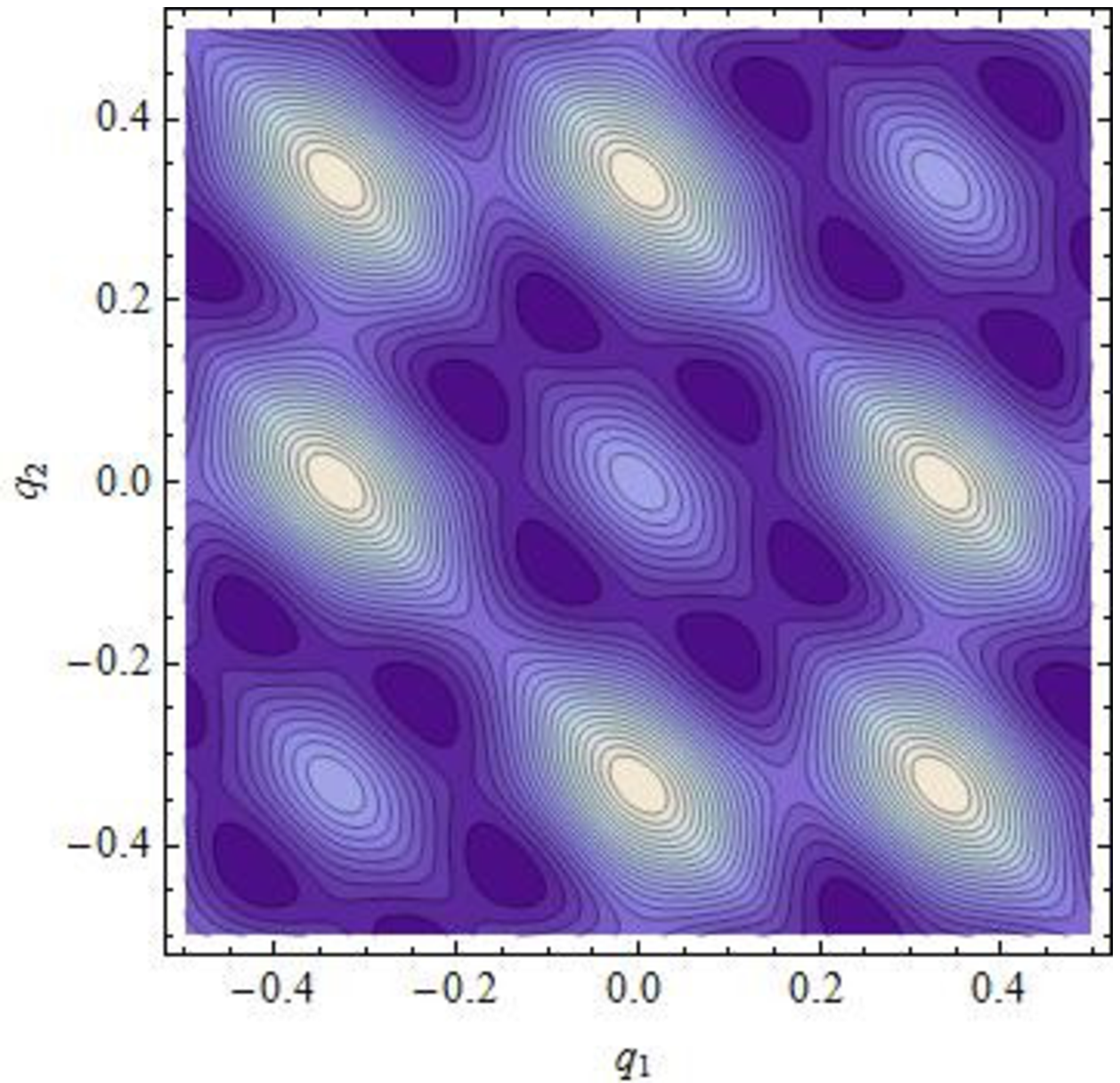}
\end{center}
\caption{
Contour plot of $[  {\cal V}_{g} + {\cal V}_{f} ] L^4$ 
in the  $q_1$-$q_2$ plane for the case of $N_{F,fund}=120$ FTBC fermions. 
The upper panel corresponds to the $SU(3)$ deconfined phase and 
the lower panel does to the $SU(2)\times U(1)$ C-broken phase.}
\label{CP}
\end{figure}

In the case of FTBC-PB,  the situation is complicated.  
If $N_{F,fund}$ is small, any nontrivial solution that induces the GB 
does not appear. 
In the case of large $N_{F,fund}\ge 30$, nontrivial solutions can survive 
at small $mL$, as shown in Fig.~\ref{CP} (bottom).  
In the case of FTBC-PB, unlike the case of ADJ-PB, 
large $N_{F,fund}$ is required for nontrivial solutions to survive. 
This difference comes from the fact that ${\cal V}_{f}(\Phi =1)-{\cal V}_{f}(\Phi =\Phi_{\rm min})$ is much smaller in the FTBC-PB case than that 
in the ADJ-PB case, where $\Phi_{\rm min}$ is a value of $\Phi$ at minimum points of ${\cal V}_f$.

For the case of FTBC-PB, the vacua corresponding to the broken phase are given by two $\mathbb{Z}_{3}$ families of solutions, 
\begin{align}
(q_{1},q_{2},q_{3})_{1}=&[(\alpha/9, \alpha/9, -2\alpha/9),
\nonumber\\
&((\alpha+3)/9,(\alpha+3)/9,(3-2\alpha)/9),
\nonumber\\
&(-(3-\alpha)/9,-(3-\alpha)/9,(6-2\alpha)/9)]
\label{V_1}
\\
(q_{1},q_{2},q_{3})_{2}=&[-(\alpha/9, \alpha/9, -2\alpha/9),
\nonumber\\
&-((\alpha+3)/9,(\alpha+3)/9,(3-2\alpha)/9),
\nonumber\\
&-(-(3-\alpha)/9,-(3-\alpha)/9,(6-2\alpha)/9)], 
\label{V_2}
\end{align}
where $\alpha$ as a function of $N_{F,fund}$ and $mL$ varies 
in a range of $0<\alpha \le1$.
In the  limit of $N_{F,fund}\to \infty$, $\alpha$ reaches $1$. 
Each family of solutions has two other choices of permutation; 
for example, $(q_1,q_2,q_3)_{1'}=(\alpha/9, -2\alpha/9, \alpha/9)$
for the first solution of (\ref{V_1}). 
In this GP phase, $SU(3)$ gauge symmetry is broken to $SU(2)\times U(1)$.

\begin{figure}[htbp]
\begin{center}
 \includegraphics[width=0.5\textwidth]{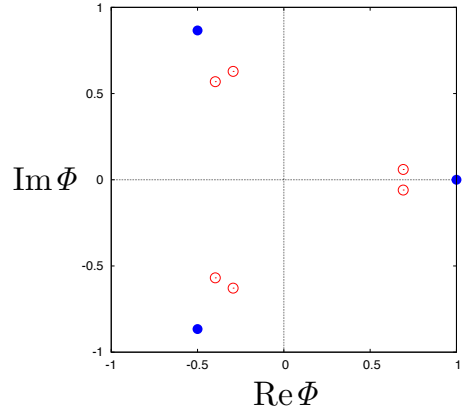}
\end{center}
\caption{
Distribution of the Polyakov loop $\Phi$ in the complex plane for a $SU(3)$ gauge theory on 
$R^{3}\times S^1$ with $N_{F,fund}=120$ FTBC fermions.
Solid circles correspond to the deconfinement phase and open circles do to the C-broken phase.
}
\label{DP}
\end{figure}

\begin{figure}[htbp]
\begin{center}
 \includegraphics[width=0.4\textwidth]{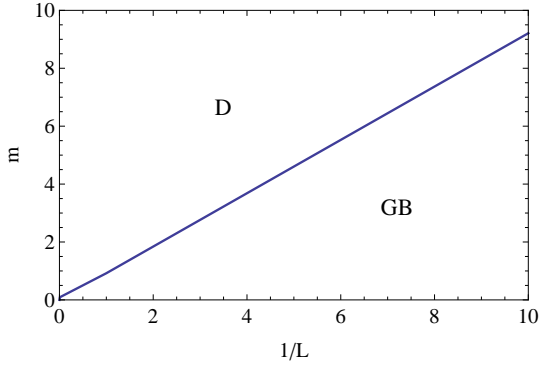}
\end{center}
\caption{The phase diagram in the $L^{-1}$-$m$ plane 
for a $SU(3)$ gauge theory on 
$R^{3}\times S^1$ with $N_{F,fund}=120$ FTBC fermions. 
The symbol D stands for the $SU(3)$ deconfinement phase 
and GB for the $SU(2)\times U(1)$ gauge-symmetry broken phase.
In the gauge-symmetry broken phase, charge conjugation is also spontaneously broken. }
\label{PD}
\end{figure}

The two families of solutions, (\ref{V_1}) and (\ref{V_2}), are related 
to each other as 
\begin{equation}
(q_{1},q_{2},q_{3})_{1}=-(q_{1},q_{2},q_{3})_{2}. 
\end{equation}
This comes from the fact that the system is invariant under charge conjugation $A_{\mu}\to -A_{\mu}$ and ${\rm Im}\Phi\to-{\rm Im} \Phi$~\cite{Kouno}.
This charge conjugation symmetry is also spontaneously broken in this 
GB phase.  
Furthermore,  $\mathbb{Z}_3$ symmetry is spontaneously violated 
in this phase, since $\Phi \neq 0$ there. 
A distribution plot of $\Phi$ is shown in Fig.~\ref{DP}, while 
the phase diagram is depicted  in Fig.~\ref{PD}.  

Here we comment on asymptotic non-freedom and renormalizability 
in a $SU(3)$ gauge theory with large $N_{F,fund}$.
In the case with $N_{F,fund}\ge 30$, the theory may lose asymptotic freedom and is expected to become non-renormalizable. 
However, the dynamical GB due 
to the Hosotani mechanism can be brought about also in the asymptotic 
non-free theory as QED or five-dimensional gauge theories. 
This is because the mechanism is based on the Aharonov-Bohm effect 
in the compact direction. 
We thus may consider that our result on the GB may be still valid, 
although we should regard our large-flavor theory on $R^{3}\times S^{1}$
as a cutoff theory.
There is no consensus on how the beta function behaves as a function of 
the number of flavors in a compactified gauge theory with special boundary conditions such as FTBC. 
Intensive study should be devoted to this topic.    

\begin{figure}[htbp]
\begin{center}
 \includegraphics[width=0.4\textwidth]{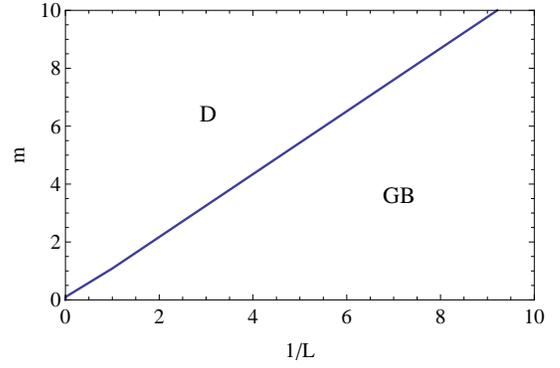}
\end{center}
\caption{The same figure as Fig. \ref{PD} but for a $SU(3)$ gauge theory on 
$R^{4}\times S^1$ with $N_{F,fund}=240$ FTBC fermions. 
}
\label{PD_5}
\end{figure}

The situation is qualitatively similar also in five-dimensional gauge theories. In this case, the effective potentials \eqref{V_gauge}$\sim$\eqref{V_ADJ} are replaced by 
\begin{eqnarray}
{\cal V}_{g}
=-{9\over{4\pi^2L^5}}\sum_{i=1}^3\sum_{j=1}^3\sum_{n=1}^{\infty}
\Bigl( 1 - \frac{1}{3} \delta_{ij} \Bigr)
\frac{\cos( 2 n \pi q_{ij})}{n^5}, 
\label{V_gauge_5}
\end{eqnarray}
\begin{eqnarray}
{\cal V}_{f,\rm FD}
&=& \frac{\sqrt{2}N_{F,fund}m^{2}}{\pi^{5/2}L^5} \sum_{i=1}^3 \sum_{n=1}^\infty
{K_{2}(nmL)\over{n^{5/2}}} 
\nonumber\\
&&\times \cos[2 \pi n (q_i+\varphi )],  
\label{V_FD_5}   
\end{eqnarray}
\begin{eqnarray}
{\cal V}_{f,\rm FTBC}
&=& \frac{\sqrt{2}N_{3}m^{2}}{L^2 \pi^{5/2}} \sum_{i=1}^3 \sum_{f=1}^3\sum_{n=1}^\infty
\nonumber\\
&&\times {K_{2}(nmL)\over{n^{5/2}}}\cos[2 \pi n (Q_{if}+\varphi )],  
\label{V_FTBC_5}   
\end{eqnarray}
and 
\begin{eqnarray}
{\cal V}_{f,\rm ADJ}
&=& \frac{\sqrt{2}N_{F,adj}m^{2}}{L^2 \pi^{5/2}} \sum_{i=1}^3 \sum_{j=1}^3\sum_{n=1}^\infty
\Bigl( 1 - \frac{1}{3} \delta_{ij} \Bigr)
\nonumber\\
&&\times {K_{2}(nmL)\over{n^{5/2}}} \cos[2 \pi n (q_{ij}+\varphi )], 
\label{V_ADJ_5}   
\end{eqnarray}
respectively.  
Figure \ref{PD_5} shows the phase diagram for a five-dimensional gauge theory with FTBC-PB fundamental fermions in the case of $N_{F,fund}=240$. 
Similarly to the four-dimensional case, the GB takes place for small $L$ and large $N_{F,fund}$. 
However, the critical flavor number is $N_{F,fund}=123$ much larger than that in the four-dimensional case. 
This difference is originated in two facts. 
The first is that the gauge degree of freedom is proportional to $d-2$ for $d$ being dimensions of spacetime, while the fermion degree of freedom is to $2^{ [d/2] }$, where the symbol ``$[~~]$" denotes the Gauss symbol.  
Therefore, the gauge degree of freedom increases more rapidly than that of fermion when $d$ increases. 
The second is more important. This is $d$ dependence of the power of $n$ 
in the denominators of ${\cal V}_f$ and ${\cal V}_g$. 
This dependence changes the range of $n$ that mostly contributes the summation of $n$ and makes ${\cal V}_g/{\cal V}_f$ larger as $d$ increases.

\section{Summary}
\label{sec:summary}

In summary,
we have investigated differences and similarities between 
fundamental and adjoint matters in $SU(N)$ gauge theories. 
The gauge theory with ordinary fundamental matter does not have 
$\mathbb{Z}_{N}$ symmetry, whereas the gauge theory 
with adjoint (ADJ) matter does. 
This implies that 
an essential difference between fundamental and ADJ matters comes 
from the presence or absence of $\mathbb{Z}_{N}$ symmetry. 
We have then imposed the FTBC on fundamental fermion in order to 
make the gauge theory $\mathbb{Z}_{N}$ symmetric, and 
have shown similarities between 
FTBC fundamental matter and ADJ matter by using the PNJL model, 
particularly for the confinement/deconfinement transition related 
to $\mathbb{Z}_{N}$ symmetry.  
Thus a main difference between ordinary fundamental matter and ADJ matter is originated in  the presence or absence of $\mathbb{Z}_{N}$ symmetry. Meanwhile, the chiral property is somewhat different 
between FTBC fundamental matter and ADJ matter, but has a simple scaling low $\sigma_{\rm FTBC}\sim \sigma_{\rm ADJ}N/N_{adj}$, 
where $\sigma_{\rm ADJ}$ ($\sigma_{\rm FTBC}$) is the chiral condensate for  ADJ (FTBC fundamental) fermion and
$N_{adj}$ ($N$) is the dimension of the adjoint (fundamental) representation of fermion and $N_{adj}$ is related to $N$ as 
$N_{adj}=N^2-1$.

We have also investigated a possibility of the gauge symmetry breaking 
(GB) at high-energy scale. 
The GB takes place for not only ADJ-PB fermion but also 
FTBC-PB fermion. 
Properties of the GB phase are different between the two fermions. 
At high-energy limit, color $SU(3)$ group is broken down to  $U(1)\times U(1)$ for ADJ-PB fermion, but to $SU(2)\times U(1)$ for FTBC-PB fermion. 
For FTBC-PB fermion, the GB phase appears only when $N_{F,fund}$ is 
large and the charge-conjugation symmetry is also  spontaneously broken there. 
The present results may suggest another class of Gauge-Higgs unification models due to the dynamical gauge symmetry breaking, 
although realization of the breaking at large $N_{F,fund}$ may mean 
that the system is not asymptotic free. 

The BCFL transformation \eqref{ZNtrans_CF} plays an important role to 
make a gauge theory $\mathbb{Z}_N$-symmetric. 
$\mathbb{Z}_N$ symmetry, i.e., invariance 
under this linked transformation is originated in 
the fact that $\mathbb{Z}_N$ group is a common subgroup of $U(1)_{\rm B}$, color $SU(N)$ and flavor $SU(N)$. 
We can classify types of diquark condensates by using the BCFL transformation. This classification is an interesting future work.

\noindent
\begin{acknowledgments}
The authors thank A. Nakamura, K. Fukushima, T. Saito and K. Nagata for valuable discussions and comments.  
H.K. also thanks M. Imachi, H. Yoneyama, H. Aoki and M. Tachibana for useful discussions. 
T.M. and K.K. appreciate the fruitful discussion with E. Itou and Y. Hosotani.
K.K. is supported by RIKEN Special Postdoctoral Researchers Program.
T.M. is supported by Grant-in-Aid for the Japan Society for Promotion of Science (JSPS) 
Postdoctoral Fellows for Research Abroad(No.24-8).
T.S. is supported by Grant-in-Aid for JSPS Fellows (No.23-2790).
\end{acknowledgments}


\end{document}